%% file: dpf2015-ds50-proc.tex
\documentclass[12pt]{article}
\pdfoutput=1
\usepackage{graphicx}
\usepackage[separate-uncertainty,retain-explicit-plus,per-mode=symbol,binary-units]{siunitx}
\usepackage[version=4]{mhchem}

\newcommand\pubnumber{DPF2015-240}
\newcommand\pubdate{\today}

\def\support{\footnote{On behalf of the DarkSide collaboration}}

\def\Title#1{\begin{center} {\Large #1 } \end{center}}
\def\Author#1{\begin{center}{ \sc #1} \end{center}}
\def\Address#1{\begin{center}{ \it #1} \end{center}}

\newcommand\pubblock{\rightline{\begin{tabular}{l} \pubnumber\\
         \pubdate  \end{tabular}}}
\newenvironment{Abstract}{\begin{quotation}  }{\end{quotation}}
\newenvironment{Presented}{\begin{quotation} \begin{center} 
             PRESENTED AT\end{center}\bigskip 	
      \begin{center}\begin{large}}{\end{large}\end{center} \end{quotation}}

\input econfmacros.tex

\DeclareSIUnit\pe{PE}
\DeclareSIUnit\yr{yr}
\newcommand\fninety{f$_{90}$}
\newcommand\arthreenine{\ce{^{39}Ar}}

\begin{document}
\begin{titlepage}
\pubblock

\vfill
\Title{Status and Results from DarkSide-50}
\vfill
\Author{ Alden Fan\support}
\Address{Physics and Astronomy Department \\ University of California, Los Angeles, CA 90095, USA}
\vfill
\begin{Abstract}
DarkSide-50 is the first physics detector of the DarkSide dark matter search program. The detector features a dual-phase underground-argon Time Projection Chamber (TPC) of 50 kg active mass surrounded by an organic liquid-scintillator neutron veto (30 tons) and a water-Cherenkov muon detector (1000 tons). The TPC is currently fully shielded and operating underground at Gran Sasso National Laboratory. A first run of 1422 kg-day exposure with atmospheric argon represents the most sensitive dark matter search using a liquid argon target. The TPC is now filled with underground argon, greatly reduced in $^{39}$Ar, and DarkSide-50 is in its final configuration for an extended dark matter search. Overviews of the design, performance, and results obtained so far with DarkSide-50 will be presented, along with future prospects for the DarkSide program.
\end{Abstract}
\vfill
\begin{Presented}
DPF 2015\\
The Meeting of the American Physical Society\\
Division of Particles and Fields\\
Ann Arbor, Michigan, August 4--8, 2015\\
\end{Presented}
\vfill
\end{titlepage}
\def\thefootnote{\fnsymbol{footnote}}
\setcounter{footnote}{0}

\section{Introduction}

DarkSide-50 is a direct dark matter detection experiment using a liquid argon (LAr) target in a dual-phase Time Projection Chamber (TPC) to detect the nuclear recoils expected from the scattering of Weakly Interacting Massive Particles (WIMPs), a favored class of dark matter particles. The DarkSide program features a number of technologies to reduce and reject electromagnetic, neutron, and surface backgrounds.

Liquid argon is an attractive target for direct dark matter detection because it is scalable to large masses, has high scintillation and ionization yields, and provides exceptional discrimination power of nuclear recoils (NRs) against electronic recoils (ERs). The dual-phase LAr TPC is sensitive to both the ionization and scintillation channels of recoiling nuclei. The scintillation light comes from the de-excitation of Ar dimers from either the singlet or triplet state, which have drastically different decay times of \SI{6}{ns} and \SI{1500}{ns}, respectively, and the ratio of singlet to triplet states differs for NR vs. ER. The time profile (pulse shape) of the scintillation light provides discrimination between the two types of recoils with high efficiency. The ratio of ionization to scintillation provides further discrimination between NR and ER.




The main challenge of using liquid argon as a WIMP detection medium is that naturally occurring atmospheric argon (AAr) has high contamination of \arthreenine, whose radioactivity limits the sensitivity of large scale detectors. The radioactive isotope has a half-life of \SI{269}{yr} and $\beta$-decays with an endpoint at \SI{565}{keV}. The isotope is cosmogenically activated, and the activity in AAr is \SI{1}{Bq/kg}. Underground argon (UAr), protected from cosmic ray activation, is significantly reduced in \arthreenine\ content. The first measurement of the UAr activity placed an upper limit on the \arthreenine\ activity at \SI{6.6}{mBq/kg} corresponding to an \arthreenine\ reduction factor of \num{>150} relative to AAr~\cite{Xu}. The discovery of underground sources of argon established the viability of argon as a direct dark matter detection medium.

This report describes the DarkSide experiment and presents results from analysis of data taken with AAr and UAr. 

\section{The DarkSide program}

DarkSide is a staged program of dark matter detectors. The first iteration, DarkSide-10 was key to the development of the dual-phase argon TPC technology, especially in developing the high voltage system and establishing an exceptionally high light yield~\cite{Alexander}. The current iteration, DarkSide-50, is the first physics-capable detector and has been operating since October 2013 deep underground at Gran Sasso National Laboratory in Italy. It is made from all low-radioactivity materials and is the first detector to use UAr. Future multi-ton-scale detectors are planned with the goal of reaching the so-called ``neutrino floor''. 

\begin{figure}[t]
\centering
\includegraphics[height=2.7in]{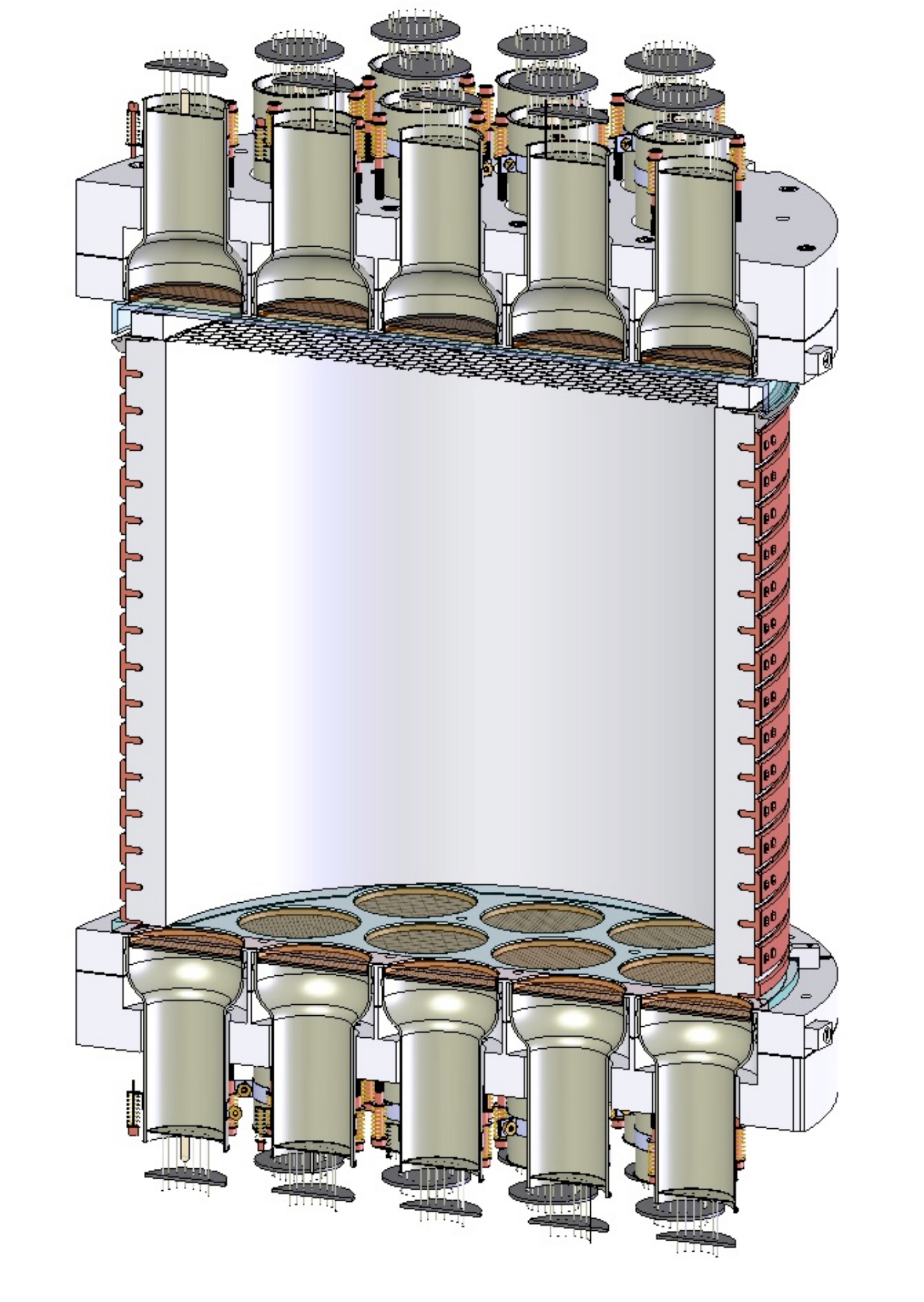}
\caption{Sketch of the DarkSide-50 TPC.}
\label{fig:tpcsketch}
\end{figure}

The DarkSide-50 experiment consists of three nested detectors. Innermost is the LAr TPC, as shown in Fig.~\ref{fig:tpcsketch}, containing the liquid argon dark matter target. Ionizing radiation enters the sensitive volume of the detector, depositing energy in the form of both scintillation and ionization. The scintillation light represents the primary pulse S1. The ionization electrons are drifted upwards by an electric field (\SI{200}{V/cm}) to the liquid surface where they are extracted into a gas pocket by a higher electric field (\SI{2.8}{kV/cm}). The acceleration of the electrons across the gas pocket produces a secondary scintillation signal S2, proportional to the number of ionization electrons. The S2 pulse enables 3D position reconstruction of the primary interaction site: the time between S1 and S2 gives the vertical position, and the hit pattern of the S2 light on the photosensors gives the transverse position. The 3D position reconstruction allows for rejection of surface backgrounds. 


 The TPC is a \SI{36}{cm} diameter by \SI{36}{cm} height cylinder with \SI{46}{kg} active volume, viewed by 38 Hamamatsu R11065 3'' PMTs. The top of the active volume is defined by a stainless steel grid, which allows setting of the drift and extraction fields independently. The gas pocket above the active volume is \SI{~1}{cm} thick. The wall of the TPC is high reflectivity PTFE and the top and bottom are fused silica windows coated with transparent conductor, ITO, forming the anode and cathode surfaces. Field shaping copper rings outside the TPC ensure a uniform electric field inside the active volume. The inner surfaces are coated with the wavelength shifter tetraphenyl butadiene (TPB) to shift the \SI{128}{\nm} scintillation light of LAr to visible \SI{420}{\nm} light. 

\begin{figure}[t]
\centering
\includegraphics[height=1.5in]{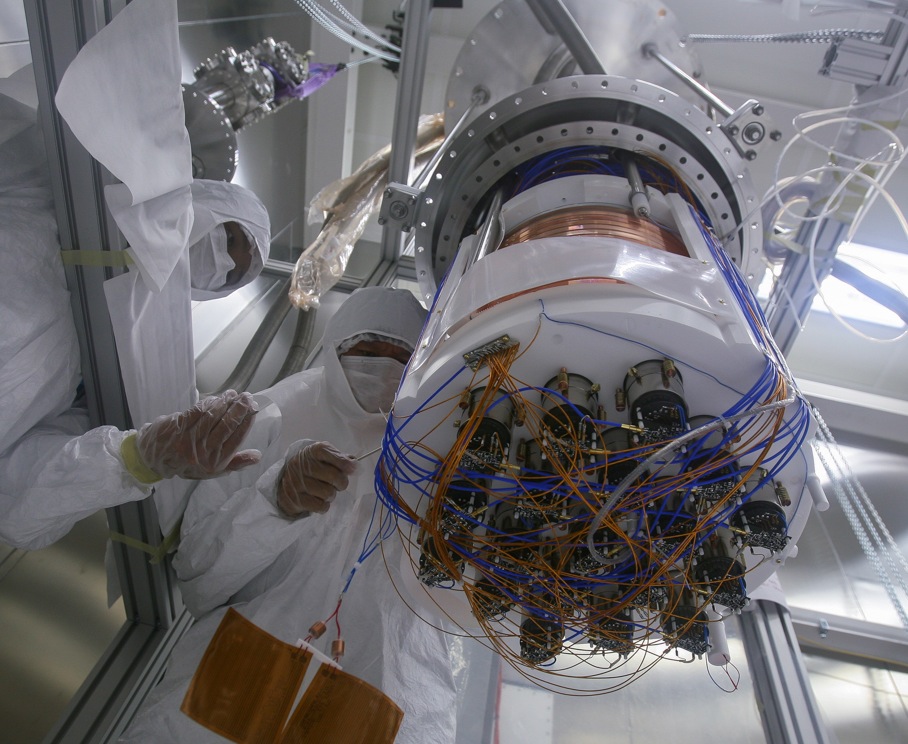}
\includegraphics[height=1.5in]{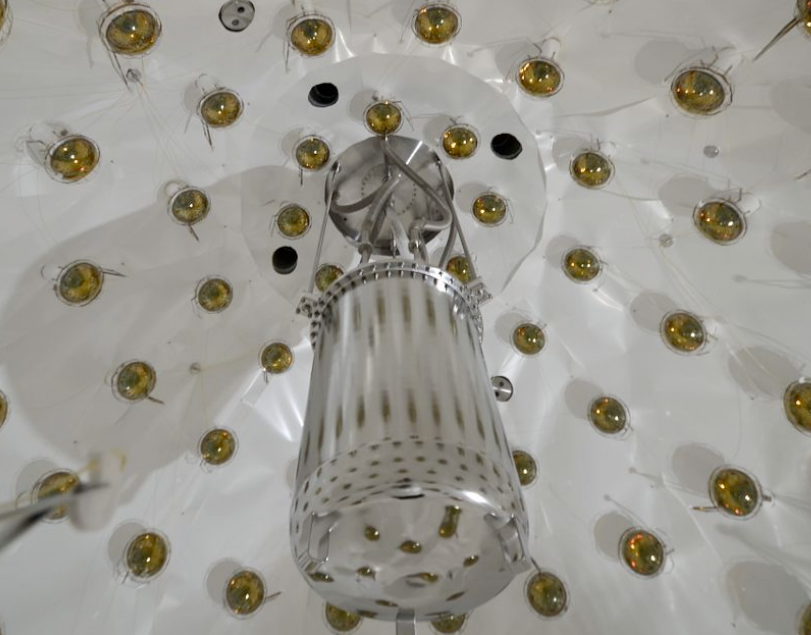}
\includegraphics[height=1.5in]{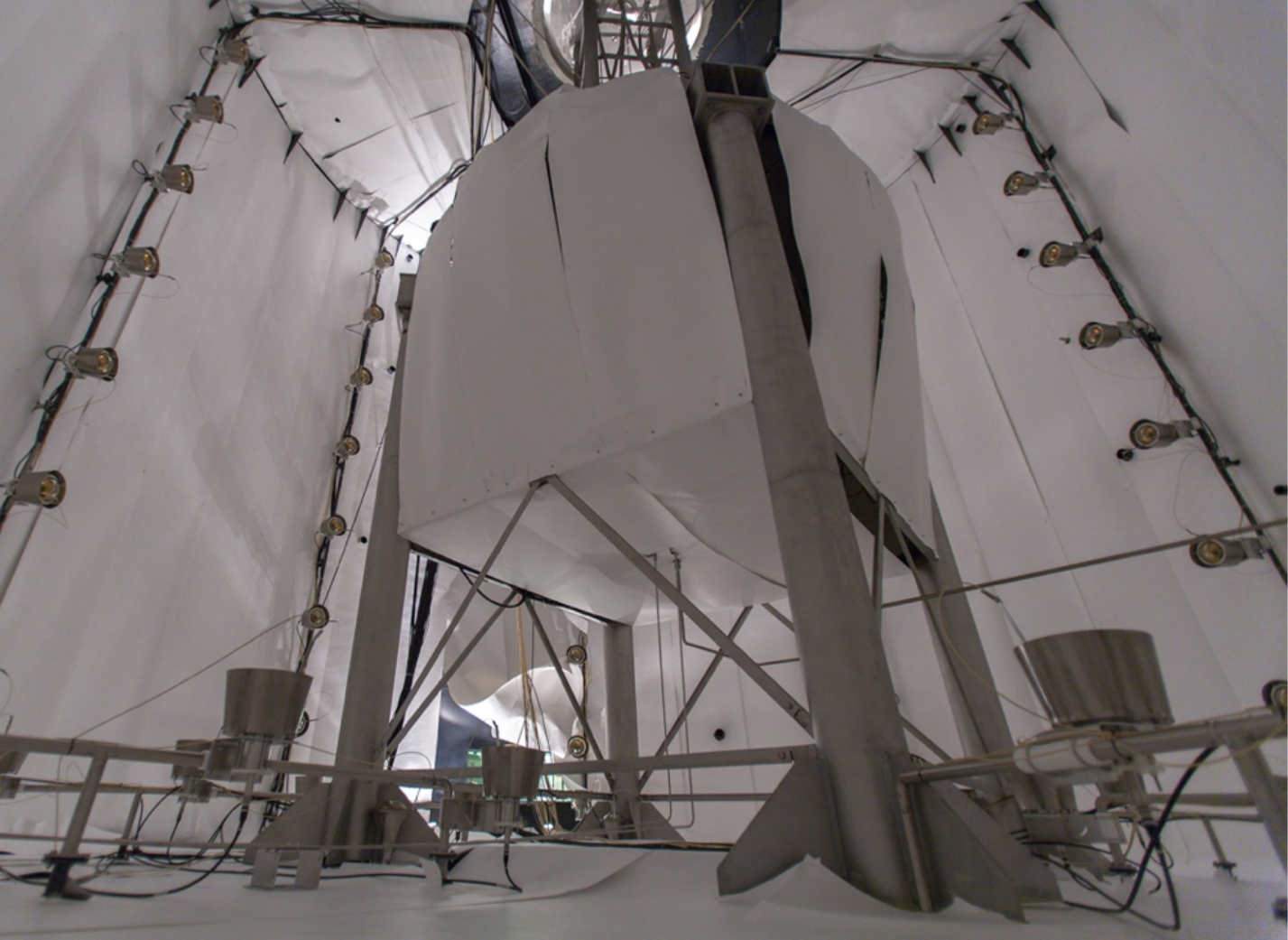}
\caption{Installation and assembly of the TPC (left), LSV (center), and WCD (right).}
\label{fig:DS50photos}
\end{figure}

The LAr TPC is surrounded by the Liquid Scintillator Veto (LSV), a \SI{4}{m} diameter sphere filled with a mixture of pseudocumene and trimethyl borate (TMB). The \ce{^{10}B} in the scintillator cocktail provides a large neutron capture cross section. The LSV is instrumented with 100 8" PMTs, making it an active veto to tag neutrons in the TPC as well as permitting \textit{in situ} measurement of the neutron background. The LSV is surrounded by the Water Cherenkov Detector (WCD), an \SI{11}{m} diameter by \SI{10}{m} height water tank, instrumented with 80 PMTs and forming an active muon veto to tag cosmogenically induced neutrons. Both the LSV and the water tank provide passive neutron and gamma shielding for the TPC. Photos from the installation and assembly of all three detectors are shown in Fig.~\ref{fig:DS50photos}.

\section{Atmospheric argon data}

The first physics results from the DarkSide-50 detector were obtained with the TPC filled with AAr~\cite{Agnes}. The data were dominated by \ce{^{39}Ar} $\beta$-decays, giving a trigger rate of \SI{~50}{Hz}. The light yield (LY) was measured \textit{in situ} by using data from a \ce{^{83m}Kr} source and by fitting the \arthreenine\ endpoint and was found to be \SI{7.9}{\pe/keV} at null field and \SI{7.0}{\pe/keV} at \SI{200}{V/cm} drift field. DarkSide-50 achieved very low levels of electronegative impurities as measured by the electron drift lifetime, which was $>$\SI{5}{ms} for the majority of the AAr campaign. The performance of the LSV was reduced due to unexpectedly high \ce{^{14}C} content in the TMB, which was found to contain modern carbon. The TMB has since been replaced with a cleaner petrolium-based sample. The LSV LY was \SI{~0.5}{\pe/keV}, measured by fitting the \ce{^{14}C} spectrum and the spectrum of \ce{^{60}Co} coming from the cryostat steel. 

The pulse shape discrimination parameter used in Ref.~\cite{Agnes} was \fninety, the fraction of S1 light collected in the first \SI{90}{ns} compared to the total S1 light, collected in a \SI{7}{\us} window. The NR \fninety\ response was extrapolated from the SCENE experiment, which exposed a small LAr TPC to a pulsed low energy neutron beam~\cite{Cao}. The NR energy scale was also transferred from SCENE to DarkSide-50 by referencing the null field LY measured with a \ce{^{83m}Kr} source in each experiment.

\begin{figure*}[t]
\centering
\includegraphics[width=0.7\textwidth]{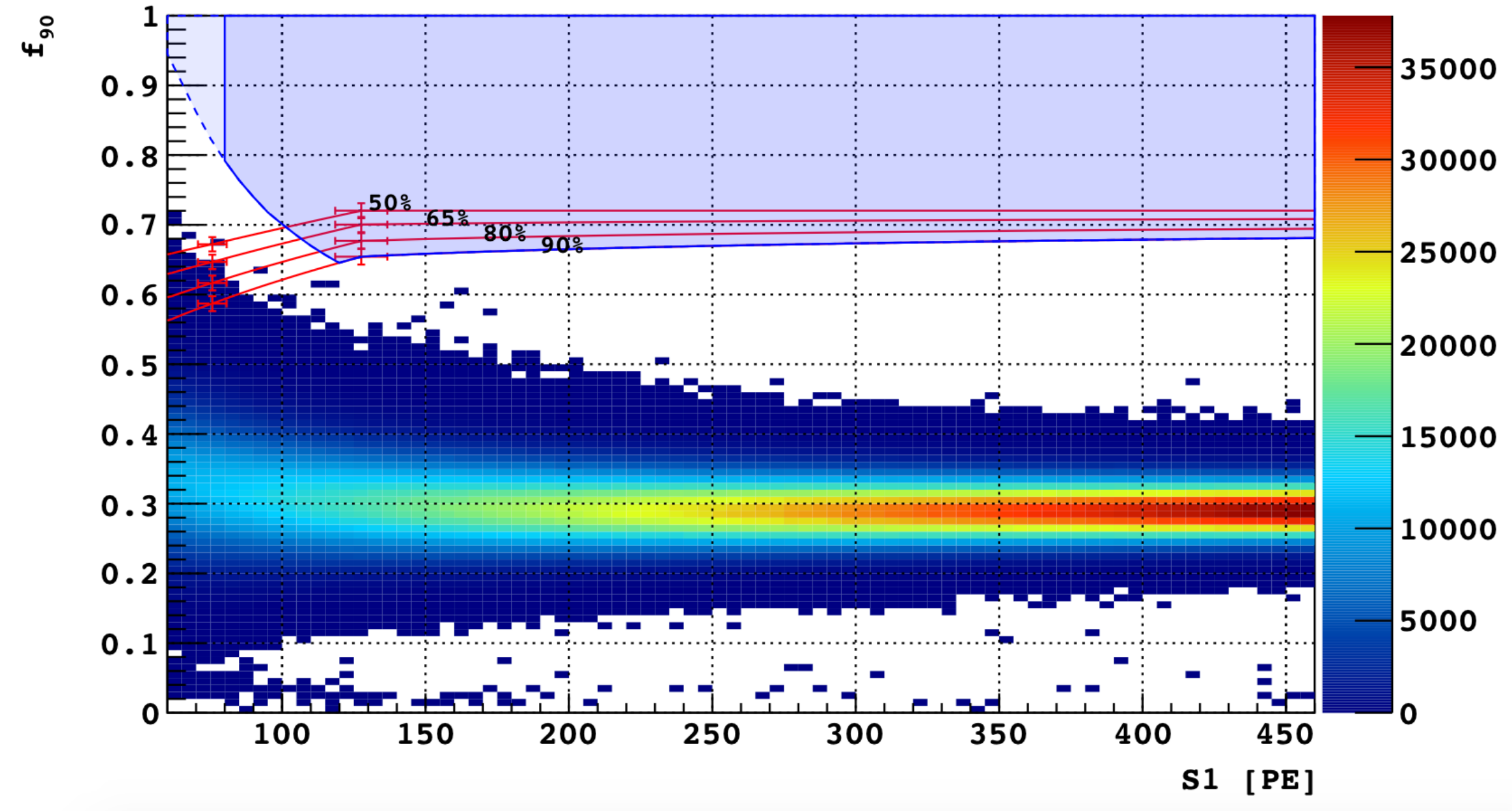}
\caption{Distribution of events in the \fninety\ vs S1 plane which survive all cuts. Shaded blue with solid blue outline: WIMP search region.  Percentages label the \fninety\ acceptance contours for NRs, drawn by connecting points at which the acceptance was determined from the corresponding SCENE measurements. } 
\label{fig:DS50AArDMsearch}
\end{figure*}

\begin{figure*}[t]
\centering
\includegraphics[width=0.7\textwidth]{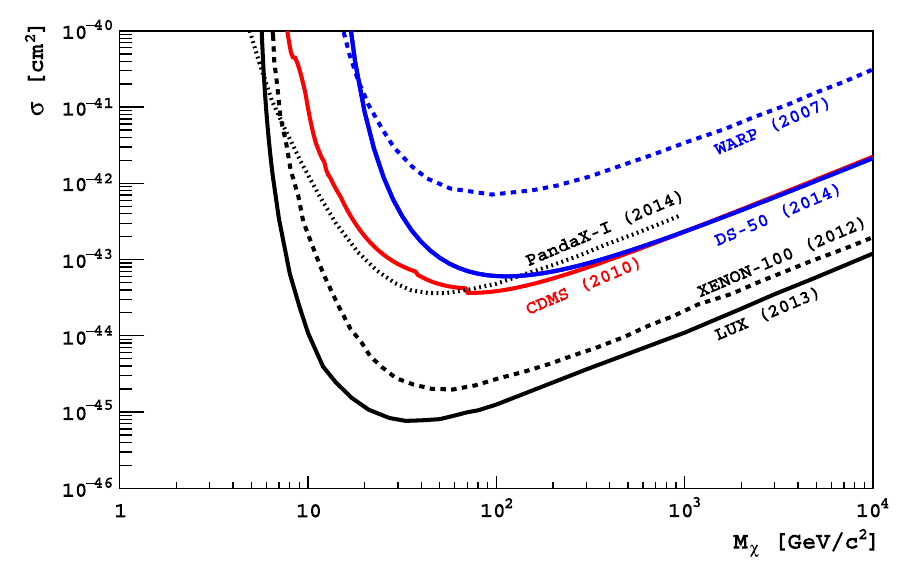}
\caption{Spin-independent WIMP-nucleon cross section 90\% C.L. exclusion plot for the DarkSide-50 AAr campaign compared to previous results from other experiments.}
\label{fig:DS50AArLimit}
\end{figure*}

The first AAr exposure contained 47.1~live days of data acquired between October 2013 and May 2014. A suite of cuts was applied to isolate single scatter recoils, removing events with \num{>1} S1 pulse or \num{>1} S2 pulse, predominantly pile-up of \arthreenine\ decays or multiple scatter $\gamma$'s; events with coincident Cherenkov light; and events with coincident signal in the LSV or the WCD. A fiducial volume cut was defined by requiring the drift time to be between \SI{40}{\us} and \SI{334.5}{\us}. No radial cuts were applied. The fiducial volume was \SI{36.9(6)}{kg}. The WIMP search region was defined in the range of \SIrange{80}{460}{\pe} of S1 (corresponding to recoils in the range \SIrange{38}{206}{keV_{\rm nr}}). The WIMP region of interest in the f$_{90}$ vs. S1 plane was defined by intersecting the 90\% nuclear recoil acceptance curve derived from SCENE with a curve corresponding to fixed \ce{^{39}Ar} leakage per S1 bin. The ER leakage was estimated using a statistical model for \fninety~\cite{Lippincott}. The total leakage of \ce{^{39}Ar} events into the WIMP box was $<$0.1~events for the \SI{1422(67)}{\kg\day} exposure. The final dark matter search, shown in Fig.~\ref{fig:DS50AArDMsearch}, resulted in zero events in the WIMP search region. This led to the placement of the 90\% CL exclusion curve shown in Fig.~\ref{fig:DS50AArLimit}, with a minimum cross section of \SI{6.1E-44}{\cm^2} at \SI{100}{GeV/c^2} WIMP mass. At the time of their publication, these results represented the most stringent WIMP dark matter search limit using a liquid argon target.

\begin{figure}[thb]
\centering
\includegraphics[width=0.8\textwidth]{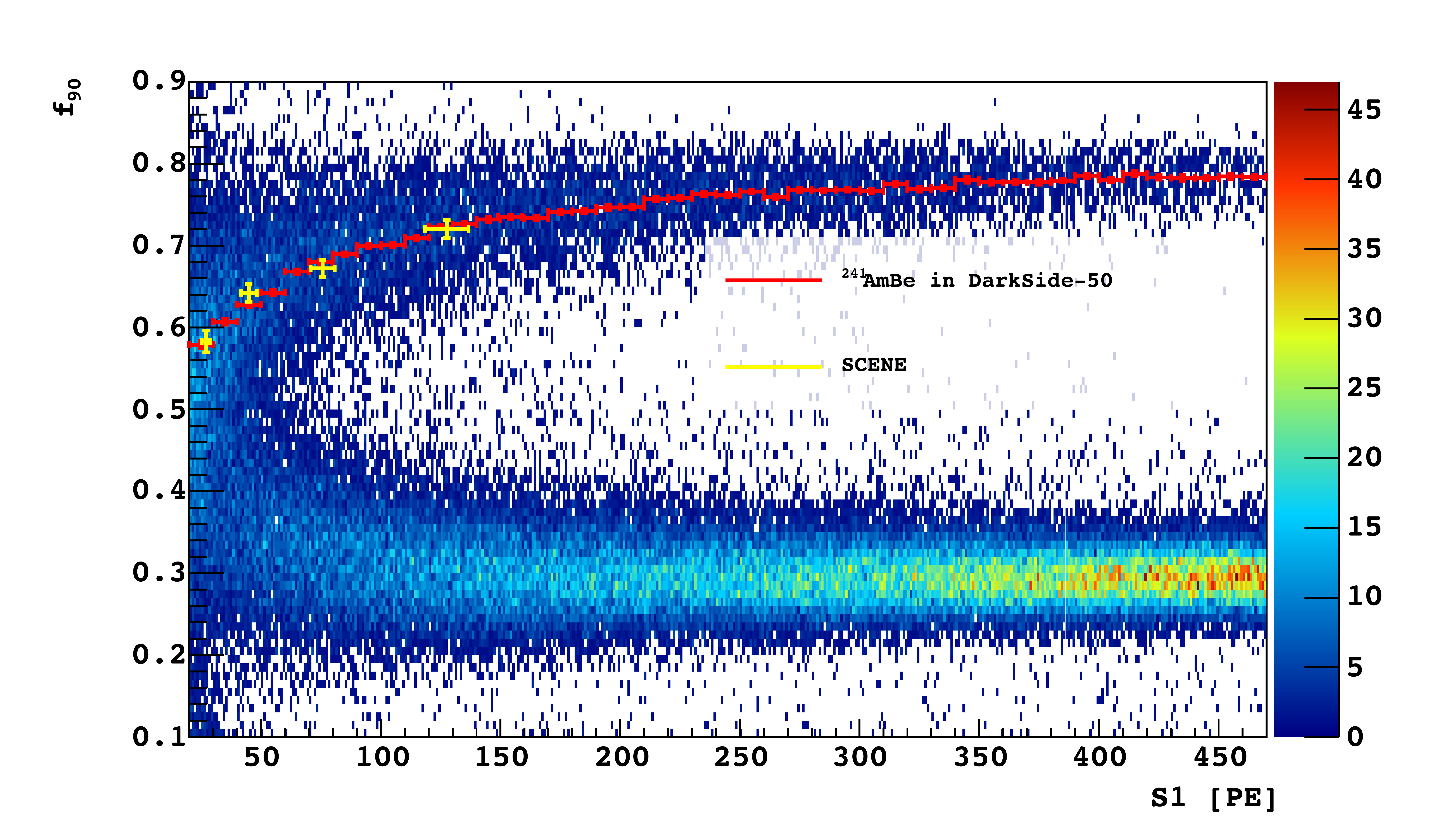}
\caption{Comparison of NR \fninety\ medians from AmBe data in DarkSide-50 (red) and extrapolated values from SCENE to DarkSide-50 (yellow). The scatter plot shows the AmBe data in DarkSide-50. Events in the region between the NR and ER bands are due to inelastic scattering of high energy neutrons, accidentals, and correlated neutron and $\gamma$-ray emission by the AmBe source. }
\label{fig:AmBe}
\end{figure}

Following the first dark matter search, the calibration hardware was installed in the DarkSide-50 apparatus in September 2014, allowing placement of a variety of gamma and neutron sources in the LSV next to the TPC cryostat. Data taken with AmBe sources validated the nuclear recoil \fninety\ response extrapolated from SCENE to DarkSide-50 as shown in Fig.~\ref{fig:AmBe}. Data taken with \ce{^{57}Co}, \ce{^{133}Ba}, and \ce{^{137}Cs} sources provided validation for DarkSide's MC simulation code. The MC code is comprehensive, including models for all three sub-detectors and models for the scintillation and recombination processes in LAr, the propagation of photons, and the electronics. The MC was tuned on the high statistics of \arthreenine\ data and validated using the various $\gamma$ calibration sources. 

\section{Underground argon data}

\begin{figure}[thb]
\centering
\includegraphics[width=0.8\textwidth]{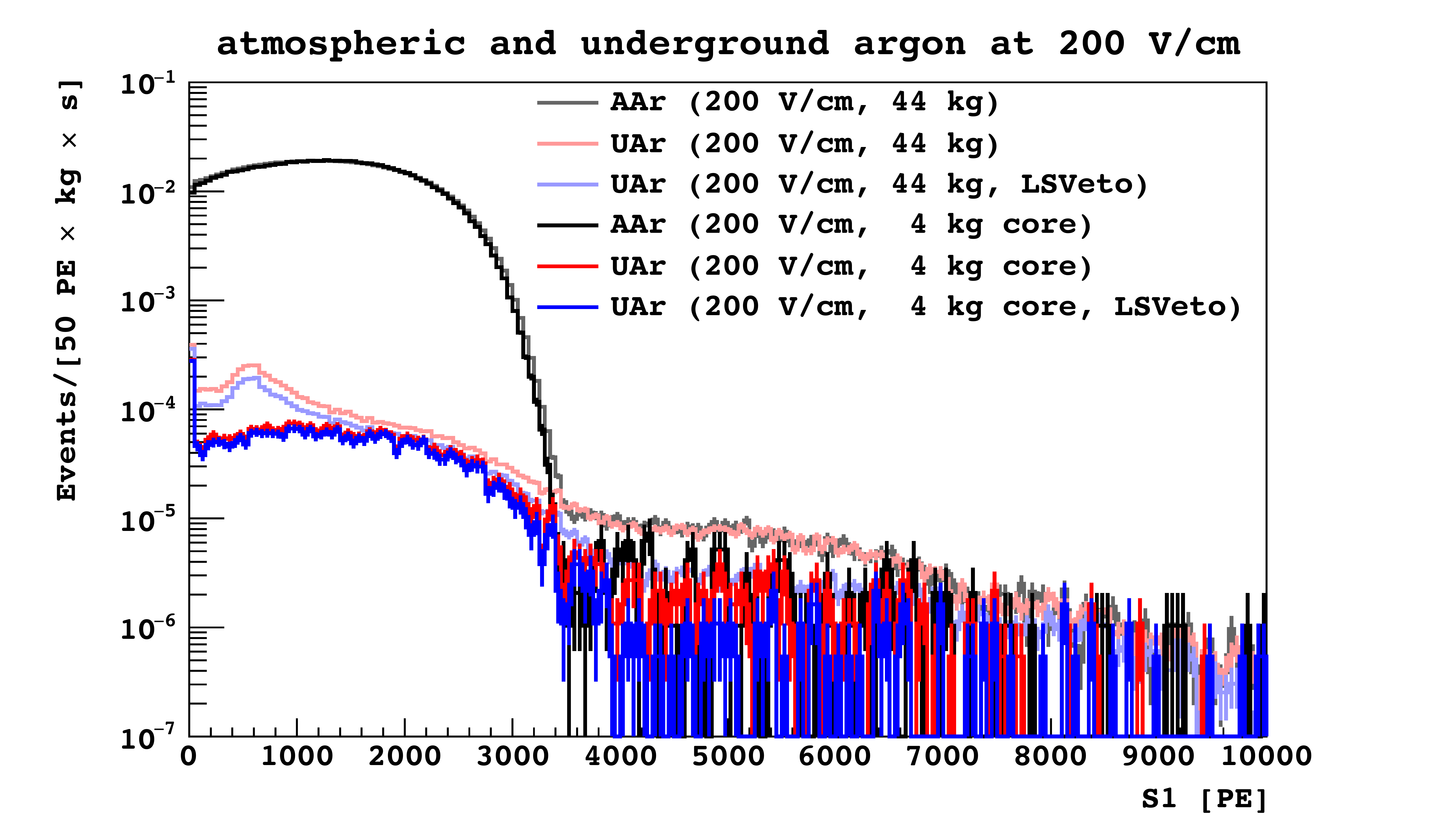}
\caption{Comparison of S1 spectra from AAr and UAr taken at \SI{200}{V/cm} drift field. Comparisons made with application of veto anti-coincidence cuts and between the full active volume and the core of the TPC. }
\label{fig:UArDepletion}
\end{figure}

The DarkSide-50 TPC was emptied of AAr in March 2015 and filled with UAr on April 1, 2015. The UAr data had a trigger rate of \SI{1.6}{Hz}, significantly reduced from the \SI{50}{Hz} rate in AAr data. With only a minimal set of cuts to select single scatter events, we saw that the rate of events beyond the \ce{^{39}Ar} endpoint agreed well between AAr and UAr, as shown in Fig.~\ref{fig:UArDepletion}, indicating that the LY was unchanged after the UAr fill. The stability of the LY was further confirmed by \ce{^{83m}Kr} calibration. An initial estimate of the reduction of \ce{^{39}Ar} activity in UAr was obtained by comparing the S1 spectra in a \SI{4}{kg} core of the TPC, where the rate of external gammas was significantly reduced. The reduction was found to be a factor of 300. That the activity from the center of the TPC was due to internal $\beta$'s was further supported by the small effect of the application of the LSV anti-coincidence cut. 

Since DPF2015, a more refined analysis of the \arthreenine\ depletion in UAr was developed using MC spectral fits~\cite{Agnes2015}. A multi-dimensional fit was performed on the null field S1 spectrum, where several background $\gamma$ lines were clearly visible; the \SI{200}{V/cm} S1 spectrum; and the drift time spectrum, which provided another handle to constrain the rates of external $\gamma$'s. The fitting procedure revealed the presence of \ce{^{85}Kr} in DarkSide-50's inventory of UAr, which was confirmed by a search for the delayed coincidences arising from the 0.43\% branching ratio to \ce{^{85m}Rb}. The fitted \arthreenine\ activity was \SI{0.73(11)}{mBq/kg} in UAr, corresponding to a depletion of \SI{1.4(2)E3} relative to AAr. 

An unexpected very low energy background component was also observed in UAr data, as can be seen in the first bin of the UAr data in Fig.~\ref{fig:UArDepletion}. This peak was identified as a \SI{2.7}{keV_{ee}} decay of \ce{^{37}Ar}, which was believed to have been cosmogenically activated while the UAr was above ground. The low energy peak was unaffected by application of the veto cuts and came from throughout the TPC volume. 
The decay of the peak was consistent with the expected \SI{35}{\day} half-life of \ce{^{37}Ar}.
Because of its short decay time, the presence of \ce{^{37}Ar} was not a concern for dark matter searches in DarkSide-50, meanwhile providing a convenient low energy calibration point for estimating the energy dependency of the TPC LY as well as estimating the trigger and reconstruction efficiencies.

\section{Conclusion}

The DarkSide-50 experiment has been operating stably since October 2013. Using atmospheric argon, the first dark matter search campaign of DarkSide-50 set the most stringent limit on the WIMP-nucleon cross section in a liquid argon target. The TPC is now filled with UAr and DarkSide-50 is prepared for an extended dark matter search. The activity of \arthreenine\ in UAr was found to be a factor 1400 lower than in AAr.

The next stage of the DarkSide program, DarkSide-20k, is a multi-ton detector with \SI{20}{\tonne} fiducial volume. It will be instrumented using silicon photomultipliers (SiPMs) which will provide an increased LY and reduced radiogenic neutron background, compared to conventional PMTs. DarkSide-20k aims for a \SI{100}{\tonne \yr} background free exposure to give a projected sensitivity of \SI{9E-48}{\cm^2} at \SI{1}{TeV/c^2}.



\end{document}

%% file: econfmacros.tex
\def\beq{\begin{equation}}
\def\eeq#1{\label{#1}\end{equation}}
\def\eeqn{\end{equation}}

\def\beqa{\begin{eqnarray}}
\def\eeqa#1{\label{#1}\end{eqnarray}}
\def\eeqan{\end{eqnarray}}

\def\Dslash{\not{\hbox{\kern-4pt $D$}}}
\def\dslash{\not{\hbox{\kern-2pt $\del$}}}

